
%
%
%
\def\unredoffs{} \def\redoffs{\voffset=-.31truein\hoffset=-.59truein}
\def\speclscape{}
%
%
%
%
\newbox\leftpage \newdimen\fullhsize \newdimen\hstitle \newdimen\hsbody
\tolerance=1000\hfuzz=2pt
\catcode`\@=11 
\def\bigans{b }
\def\answ{b }

\ifx\answ\bigans\message{(This will come out unreduced.}
\magnification=1200\unredoffs\baselineskip=16pt plus 2pt minus 1pt
\hsbody=\hsize \hstitle=\hsize 
\else\message{(This will be reduced.} \let\l@r=L
\magnification=1000\baselineskip=16pt plus 2pt minus 1pt \vsize=7truein
\redoffs \hstitle=8truein\hsbody=4.75truein\fullhsize=10truein\hsize=\hsbody
\output={\ifnum\pageno=0 
  \shipout\vbox{\speclscape{\hsize\fullhsize\makeheadline}
   \hbox to \fullhsize{\hfill\pagebody\hfill}}\advancepageno
  \else
 \almostshipout{\leftline{\vbox{\pagebody\makefootline}}}\advancepageno
  \fi}
\def\almostshipout#1{\if L\l@r \count1=1 \message{[\the\count0.\the\count1]}
      \global\setbox\leftpage=#1 \global\let\l@r=R
 \else \count1=2
  \shipout\vbox{\speclscape{\hsize\fullhsize\makeheadline}
      \hbox to\fullhsize{\box\leftpage\hfil#1}}  \global\let\l@r=L\fi}
\fi
%
\newcount\yearltd\yearltd=\year\advance\yearltd by -1900

\def\Title#1#2{\nopagenumbers\abstractfont\hsize=\hstitle\rightline{#1}%
\vskip 1in\centerline{\titlefont #2}\abstractfont\vskip .5in\pageno=0}
\def\Date#1{\vfill\leftline{#1}\tenpoint\supereject\global\hsize=\hsbody%
\footline={\hss\tenrm\folio\hss}}
%

\def\draftmode{\message{ DRAFTMODE }\def\draftdate{{\rm preliminary draft:
\number\month/\number\day/\number\yearltd\ \ \hourmin}}%
\headline={\hfil\draftdate}\writelabels\baselineskip=20pt plus 2pt minus 2pt
 {\count255=\time\divide\count255 by 60 \xdef\hourmin{\number\count255}
  \multiply\count255 by-60\advance\count255 by\time
  \xdef\hourmin{\hourmin:\ifnum\count255<10 0\fi\the\count255}}}
\def\nolabels{\def\wrlabeL##1{}\def\eqlabeL##1{}\def\reflabeL##1{}}
\def\writelabels{\def\wrlabeL##1{\leavevmode\vadjust{\rlap{\smash%
{\line{{\escapechar=` \hfill\rlap{\sevenrm\hskip.03in\string##1}}}}}}}%
\def\eqlabeL##1{{\escapechar-1\rlap{\sevenrm\hskip.05in\string##1}}}%
\def\reflabeL##1{\noexpand\llap{\noexpand\sevenrm\string\string\string##1}}}
\nolabels
%
\global\newcount\secno \global\secno=0
\global\newcount\meqno \global\meqno=1
\def\newsec#1{\global\advance\secno by1\message{(\the\secno. #1)}
\global\subsecno=0\eqnres@t\noindent{\bf\the\secno. #1}
\writetoca{{\secsym} {#1}}\par\nobreak\medskip\nobreak}
\def\eqnres@t{\xdef\secsym{\the\secno.}\global\meqno=1\bigbreak\bigskip}
\def\sequentialequations{\def\eqnres@t{\bigbreak}}\xdef\secsym{}
\global\newcount\subsecno \global\subsecno=0
\def\subsec#1{\global\advance\subsecno by1\message{(\secsym\the\subsecno. #1)}
\ifnum\lastpenalty>9000\else\bigbreak\fi
\noindent{\it\secsym\the\subsecno. #1}\writetoca{\string\quad
{\secsym\the\subsecno.} {#1}}\par\nobreak\medskip\nobreak}
\def\appendix#1#2{\global\meqno=1\global\subsecno=0\xdef\secsym{\hbox{#1.}}
\bigbreak\bigskip\noindent{\bf Appendix #1. #2}\message{(#1. #2)}
\writetoca{Appendix {#1.} {#2}}\par\nobreak\medskip\nobreak}
%
%
\def\eqnn#1{\xdef #1{(\secsym\the\meqno)}\writedef{#1\leftbracket#1}%
\global\advance\meqno by1\wrlabeL#1}
\def\eqna#1{\xdef #1##1{\hbox{$(\secsym\the\meqno##1)$}}
\writedef{#1\numbersign1\leftbracket#1{\numbersign1}}%
\global\advance\meqno by1\wrlabeL{#1$\{\}$}}
\def\eqn#1#2{\xdef #1{(\secsym\the\meqno)}\writedef{#1\leftbracket#1}%
\global\advance\meqno by1$$#2\eqno#1\eqlabeL#1$$}
%
\newskip\footskip\footskip14pt plus 1pt minus 1pt 
\def\footnotefont{\ninepoint}\def\f@t#1{\footnotefont #1\@foot}
\def\f@@t{\baselineskip\footskip\bgroup\footnotefont\aftergroup\@foot\let\next}
\setbox\strutbox=\hbox{\vrule height9.5pt depth4.5pt width0pt}
\global\newcount\ftno \global\ftno=0
\def\foot{\global\advance\ftno by1\footnote{$^{\the\ftno}$}}
%
\newwrite\ftfile
\def\footend{\def\foot{\global\advance\ftno by1\chardef\wfile=\ftfile
$^{\the\ftno}$\ifnum\ftno=1\immediate\openout\ftfile=foots.tmp\fi%
\immediate\write\ftfile{\noexpand\smallskip%
\noexpand\item{f\the\ftno:\ }\pctsign}\findarg}%
\def\footatend{\vfill\eject\immediate\closeout\ftfile{\parindent=20pt
\centerline{\bf Footnotes}\nobreak\bigskip\input foots.tmp }}}
\def\footatend{}
%
%
\global\newcount\refno \global\refno=1
\newwrite\rfile
\def\ref{[\the\refno]\nref}
\def\nref#1{\xdef#1{[\the\refno]}\writedef{#1\leftbracket#1}%
\ifnum\refno=1\immediate\openout\rfile=refs.tmp\fi
\global\advance\refno by1\chardef\wfile=\rfile\immediate
\write\rfile{\noexpand\item{#1\ }\reflabeL{#1\hskip.31in}\pctsign}\findarg}
\def\findarg#1#{\begingroup\obeylines\newlinechar=`\^^M\pass@rg}
{\obeylines\gdef\pass@rg#1{\writ@line\relax #1^^M\hbox{}^^M}%
\gdef\writ@line#1^^M{\expandafter\toks0\expandafter{\striprel@x #1}%
\edef\next{\the\toks0}\ifx\next\em@rk\let\next=\endgroup\else\ifx\next\empty%
\else\immediate\write\wfile{\the\toks0}\fi\let\next=\writ@line\fi\next\relax}}
\def\striprel@x#1{} \def\em@rk{\hbox{}}
\def\lref{\begingroup\obeylines\lr@f}
\def\lr@f#1#2{\gdef#1{\ref#1{#2}}\endgroup\unskip}

\def\addref#1{\immediate\write\rfile{\noexpand\item{}#1}} 
\def\startrefs#1{\immediate\openout\rfile=refs.tmp\refno=#1}
\def\xref{\expandafter\xr@f}\def\xr@f[#1]{#1}
\def\refs#1{\count255=1[\r@fs #1{\hbox{}}]}
\def\r@fs#1{\ifx\und@fined#1\message{reflabel \string#1 is undefined.}%
\nref#1{need to supply reference \string#1.}\fi%
\vphantom{\hphantom{#1}}\edef\next{#1}\ifx\next\em@rk\def\next{}%
\else\ifx\next#1\ifodd\count255\relax\xref#1\count255=0\fi%
\else#1\count255=1\fi\let\next=\r@fs\fi\next}
%

%
\newwrite\ffile\global\newcount\figno \global\figno=1
\def\fig{fig.~\the\figno\nfig}
\def\nfig#1{\xdef#1{fig.~\the\figno}%
\writedef{#1\leftbracket fig.\noexpand~\the\figno}%
\ifnum\figno=1\immediate\openout\ffile=figs.tmp\fi\chardef\wfile=\ffile%
\immediate\write\ffile{\noexpand\medskip\noexpand\item{Fig.\ \the\figno. }
\reflabeL{#1\hskip.55in}\pctsign}\global\advance\figno by1\findarg}
\def\xfig{\expandafter\xf@g}\def\xf@g fig.\penalty\@M\ {}
\def\figs#1{figs.~\f@gs #1{\hbox{}}}
\def\f@gs#1{\edef\next{#1}\ifx\next\em@rk\def\next{}\else
\ifx\next#1\xfig #1\else#1\fi\let\next=\f@gs\fi\next}
\newwrite\lfile
{\escapechar-1\xdef\pctsign{\string\%}\xdef\leftbracket{\string\{}
\xdef\rightbracket{\string\}}\xdef\numbersign{\string\#}}

\def\writestop{\def\writestoppt{\immediate\write\lfile{\string\pageno%
\the\pageno\string\startrefs\leftbracket\the\refno\rightbracket%
\string\def\string\secsym\leftbracket\secsym\rightbracket%
\string\secno\the\secno\string\meqno\the\meqno}\immediate\closeout\lfile}}
\def\writestoppt{}\def\writedef#1{}
\def\seclab#1{\xdef #1{\the\secno}\writedef{#1\leftbracket#1}\wrlabeL{#1=#1}}
\def\subseclab#1{\xdef #1{\secsym\the\subsecno}%
\writedef{#1\leftbracket#1}\wrlabeL{#1=#1}}
\newwrite\tfile \def\writetoca#1{}
\def\leaderfill{\leaders\hbox to 1em{\hss.\hss}\hfill}
\def\writetoc{\immediate\openout\tfile=toc.tmp
   \def\writetoca##1{{\edef\next{\write\tfile{\noindent ##1
   \string\leaderfill {\noexpand\number\pageno} \par}}\next}}}

%
\catcode`\@=12 
%
\edef\tfontsize{\ifx\answ\bigans scaled\magstep3\else scaled\magstep4\fi}
\font\titlerm=cmr10 \tfontsize \font\titlerms=cmr7 \tfontsize
\font\titlermss=cmr5 \tfontsize \font\titlei=cmmi10 \tfontsize
\font\titleis=cmmi7 \tfontsize \font\titleiss=cmmi5 \tfontsize
\font\titlesy=cmsy10 \tfontsize \font\titlesys=cmsy7 \tfontsize
\font\titlesyss=cmsy5 \tfontsize \font\titleit=cmti10 \tfontsize
\skewchar\titlei='177 \skewchar\titleis='177 \skewchar\titleiss='177
\skewchar\titlesy='60 \skewchar\titlesys='60 \skewchar\titlesyss='60
\def\titlefont{\def\rm{\fam0\titlerm}
\textfont0=\titlerm \scriptfont0=\titlerms \scriptscriptfont0=\titlermss
\textfont1=\titlei \scriptfont1=\titleis \scriptscriptfont1=\titleiss
\textfont2=\titlesy \scriptfont2=\titlesys \scriptscriptfont2=\titlesyss
\textfont\itfam=\titleit \def\it{\fam\itfam\titleit}\rm}
 \ifx\answ\bigans\else scaled\magstep1\fi
\ifx\answ\bigans\def\abstractfont{\tenpoint}\else
\font\abssl=cmsl10 scaled \magstep1
\font\absrm=cmr10 scaled\magstep1 \font\absrms=cmr7 scaled\magstep1
\font\absrmss=cmr5 scaled\magstep1 \font\absi=cmmi10 scaled\magstep1
\font\absis=cmmi7 scaled\magstep1 \font\absiss=cmmi5 scaled\magstep1
\font\abssy=cmsy10 scaled\magstep1 \font\abssys=cmsy7 scaled\magstep1
\font\abssyss=cmsy5 scaled\magstep1 \font\absbf=cmbx10 scaled\magstep1
\skewchar\absi='177 \skewchar\absis='177 \skewchar\absiss='177
\skewchar\abssy='60 \skewchar\abssys='60 \skewchar\abssyss='60
\def\abstractfont{\def\rm{\fam0\absrm}
\textfont0=\absrm \scriptfont0=\absrms \scriptscriptfont0=\absrmss
\textfont1=\absi \scriptfont1=\absis \scriptscriptfont1=\absiss
\textfont2=\abssy \scriptfont2=\abssys \scriptscriptfont2=\abssyss
\textfont\itfam=\bigit \def\it{\fam\itfam\bigit}\def\footnotefont{\tenpoint}%
\textfont\slfam=\abssl \def\sl{\fam\slfam\abssl}%
\textfont\bffam=\absbf \def\bf{\fam\bffam\absbf}\rm}\fi
\def\tenpoint{\def\rm{\fam0\tenrm}
\textfont0=\tenrm \scriptfont0=\sevenrm \scriptscriptfont0=\fiverm
\textfont1=\teni  \scriptfont1=\seveni  \scriptscriptfont1=\fivei
\textfont2=\tensy \scriptfont2=\sevensy \scriptscriptfont2=\fivesy
\textfont\itfam=\tenit \def\it{\fam\itfam\tenit}\def\footnotefont{\ninepoint}%
\textfont\bffam=\tenbf \def\bf{\fam\bffam\tenbf}\def\sl{\fam\slfam\tensl}\rm}
\font\ninerm=cmr9 \font\sixrm=cmr6 \font\ninei=cmmi9 \font\sixi=cmmi6
\font\ninesy=cmsy9 \font\sixsy=cmsy6 \font\ninebf=cmbx9
\font\nineit=cmti9 \font\ninesl=cmsl9 \skewchar\ninei='177
\skewchar\sixi='177 \skewchar\ninesy='60 \skewchar\sixsy='60
\def\ninepoint{\def\rm{\fam0\ninerm}
\textfont0=\ninerm \scriptfont0=\sixrm \scriptscriptfont0=\fiverm
\textfont1=\ninei \scriptfont1=\sixi \scriptscriptfont1=\fivei
\textfont2=\ninesy \scriptfont2=\sixsy \scriptscriptfont2=\fivesy
\textfont\itfam=\ninei \def\it{\fam\itfam\nineit}\def\sl{\fam\slfam\ninesl}%
\textfont\bffam=\ninebf \def\bf{\fam\bffam\ninebf}\rm}
%
%

\hyphenation{anom-aly anom-alies coun-ter-term coun-ter-terms}
\def\inv{^{\raise.15ex\hbox{${\scriptscriptstyle -}$}\kern-.05em 1}}

\def\Dsl{\,\raise.15ex\hbox{/}\mkern-13.5mu D} 
\def\dsl{\raise.15ex\hbox{/}\kern-.57em\partial}

\font\bigit=cmti10 scaled \magstep1
\def\lspace{\ifx\answ\bigans{}\else\qquad\fi}
\def\lbspace{\ifx\answ\bigans{}\else\hskip-.2in\fi} 
\def\boxeqn#1{\vcenter{\vbox{\hrule\hbox{\vrule\kern3pt\vbox{\kern3pt
	\hbox{${\displaystyle #1}$}\kern3pt}\kern3pt\vrule}\hrule}}}
\def\mbox#1#2{\vcenter{\hrule \hbox{\vrule height#2in
		\kern#1in \vrule} \hrule}}  
%

\def\darr#1{\raise1.5ex\hbox{$\leftrightarrow$}\mkern-16.5mu #1}

\def\half{{\textstyle{1\over2}}} 
\def\roughly#1{\raise.3ex\hbox{$#1$\kern-.75em\lower1ex\hbox{$\sim$}}}


\def\half{{1\over 2}}
\def\der#1{{d{#1}(\ell)\over d\ell}}
\def\nml{\hbox{$\cal N$}}

\def\fint#1{\int {d{#1}\over 2\pi}}

\def\rsq#1{\hbox{${\langle\,[ r({#1}) - r(0) ]^2\,\rangle}$}}

\Title{IASSNS-HEP-93/19}{\vbox{\centerline{Flory Exponents from a
Self-Consistent}
\vskip2pt\centerline{Renormalization Group}}}

\centerline{Randall D. Kamien\foot{email: kamien@guinness.ias.edu}}
\bigskip
\centerline{School of Natural Sciences}
\centerline{Institute for Advanced Study}
\centerline{Princeton, NJ 08540}

\vskip .3in
The wandering exponent $\nu$ for an isotropic polymer is predicted
remarkably well by a simple argument due to Flory.
By considering oriented polymers living in a one-parameter family of
background tangent fields, we are able to relate the wandering exponent to
the exponent in the background field through an $\epsilon$-expansion.
We then choose the background field
to have the same correlations as the individual polymer, thus self-consistently
solving for $\nu$.
We find $\nu=3/(d+2)$ for $d< 4$ and $\nu=1/2$ for $d\ge 4$,
which is exactly the Flory result.
\Date{2 April 1993; Revised 2 June 1993}

\newsec{Introduction}

The theory of polymer conformations is at once elegant and confounding.  Flory
theory, which is based on dimensional analysis, does a remarkable job in
predicting the wandering exponent $\nu$ defined by $\rsq{L}\sim L^{2\nu}$
\ref\FLOR
{P.~Flory, Principles of Polymer Chemistry, Chap. XII, (Cornell
University Press, Ithaca, 1971).}.  Nonetheless,
controlled approximations which attempt to improve on Flory theory for polymers
are not nearly as good.
Flory theory predicts $\nu_F=1/2$ for $d\ge 4$ and $\nu_F=3/(d+2)$ for $d\le
4$,
which is exact in and above the
critical dimension $d_c=4$, where the polymer acts as a random walk \ref\bou
{For another way of interpreting the Flory exponent, see
J.~P.~Bouchaud and A.~Georges, Physics Reports {\bf 195} (1990) 127, and
references therein.}.
It has been shown \ref\saleur{B. Nienhuis, Phys. Rev. Lett. {\bf 49} (1982)
1062.}\
that $\nu_F$ is exact
in $d=2$ dimensions as well.
In three-dimensions
the resummed $\epsilon$-expansion gives $\nu=0.5880\pm 0.0010$
\ref\lzj{P.G.~de~Gennes, Phys. Lett. A {\bf 44} (1973) 271; J.C.~Le~Guillou
and J.~Zinn-Justin, Phys. Rev. Lett. {\bf 39} (1977) 95.},
Flory theory gives $\nu_F=0.6$, and the most recent numerical simulations
cannot
distinguish between the two \ref\gre{G.~Grest, private communication.}.

In this note we rederive the Flory exponent.  We do this by considering
oriented
polymers interacting with a background directing field.  Flory theory
would predict the same value of $\nu$ for oriented or non-oriented polymers.
For a family of directing
fields, parameterized by $\Delta$, we can derive an exact result for $\nu$ in
terms of
$\Delta$.  Unlike other self-consistent analyses,
we find the self-consistent exponent by
matching exponents, not correlation function
prefactors \ref\bouii{J.~P.~Bouchaud and M.~E.~Cates,
Phys. Rev. E {\bf 47} (1993) 1455; J.~Doherty, M.~A.~Moore and A.~Bray
(unpublished).}.
We then choose $\Delta$ so that the tangent-tangent correlation function
of the polymer scales with the same exponent as the background field
correlation, thus
self-consistently choosing $\Delta$.  We find the Flory value, namely $\nu=1/z
= 3/(d+2)$,
where $z$ is the dynamical exponent, as we discuss below.  It is amusing that
the Flory
result comes from an $\epsilon$ expansion which happens to be, in this case,
exact.  This
may suggest why Flory theory is so good.  Additionally, it suggests how one
might
study tethered surfaces, where Flory theory is not so good and
$\epsilon$-expansions
are not so easy.

\newsec{Formulation}

We consider a $d$ dimensional system with a background
tangent field $ u$,
\eqn\erandf{Z_u[r,s] = \nml {\int}^{r(s)=r}_{r(0)=0} [dr]
\exp\left[ -{1\over 4D}\int_0^s ds'\,\left( {dr(s')\over ds'} - \lambda
u(r(s'),s')\right)^2\right],}
where $s$ labels the monomer along the polymer.
Then the annealed partition function is
\eqn\zann{Z_{\hbox{an}}(r,s) = \int [du] Z_u(r,s) P[u].}
A factor of $\lambda$ has been
introduced to help
to organize the perturbation expansion, and will, in the end, be set to unity.
In addition, it is needed to make the units correct, i.e.  $[\lambda ] =
LS^{-1}$.

The vector field ${u}({x},s)$ is a function of both space {\sl and}
the monomer label $s$, as if each monomer interacts with a different vector
field, even though they could be near each other in space.  Note that in mean
field
theory
${dr(s)/ds} = \lambda u(r(s),s)$.  Consider the tangent correlations of two
monomers at two nearby points in space, $\cal A$ and $\cal B$.  If the polymer
takes
a short route from $\cal A$ to $\cal B$, then the tangent vectors along that
length
will be strongly correlated, since they must mostly lie along the line
connecting
the two points.  On the other hand, if the polymer takes a long circuitous path
while
getting from $\cal A$ to $\cal B$, the two tangent vectors will not be very
correlated.
Since $u(x,s)$ is the self-consistent tangent field, it must reflect this
simple geometric
argument.  If $u$ did {\sl not} depend on $s$, the tangent-tangent correlations
along
the polymer would only depend on their distance in space, and would not respect
the constraint relating the path to the tangent vectors.

We now view the functional integral over $r(s)$ as a quantum mechanical
propagator in imaginary time.
That is, $Z_u$ will satisfy a Fokker-Planck equation with the initial
condition $r(0)=0$.  In this case the Euclidean Lagrangian is just our
Hamiltonian.  The Euclidean Hamiltonian is the Legendre
transform of the Lagrangian.  However, as with the theory of a point particle
in an electromagnetic field, there is an ordering ambiguity.  The
functional integral produces the symmetric Hamiltonian.
Because, in statistical mechanics, the Hamiltonian comes from a transfer
matrix,
we must take the symmetric ordering of the momentum and the velocity
field, as is done in electrodynamics of a quantum mechanical particle
\ref\rFEYN{R.P. Feynman, {Rev.\ Mod.\ Phys.\ }{\bf 20} (1948) 367.}.

The Fokker-Planck (Euclidean Schr\"odinger) equation obtained is:
\eqn\erwh{{\partial Z_u(r,s)\over \partial s} = D\nabla^2Z_u(r,s) -
\half\lambda\left\{\nabla\!\cdot\!\left[ u(r,s)Z_u(r,s)\right] +
u(r,s)\cdot\nabla
Z_u(r,s)\right\}.}
If we write $\psi (r,s)=\theta (s)Z_u(r,s)$ and Fourier transform in space and
time, we get:
\eqn\ething{(-i\omega + Dk^2)\psi (k,\omega ) = \psi_0(k) - \half (i\lambda )
\fint{\eta}\mint{q}\,u_i(k-q)(k^i+q^i)\psi (q,\eta )}
where $\omega$ is the fourier variable conjugate to $s$, $k$ is the
fourier variable conjugate to $x$ and $\psi_0$ is the
fourier transform of the initial conditions.  The
boundary condition $r(0)=0$ corresponds to $\psi_0 \equiv 1$.  Equation
\ething\ can be solved recursively in powers of
$\lambda$.

Now we must consider the random field $u$.  If $u$ is to describe
the self-consistent background, then, if the polymers have no ends (by either
being cyclic or spanning the system), $\nabla\!\cdot\! u=0$.  With this
constraint,
we take
\eqn\euu{\langle\,u_i(k,\omega)u_j(k',\omega')\,\rangle
= {\delta_{ij} - {k_ik_j/k^2}\over A\omega^2 +
Bk^{2\Delta}}\,\delta(\omega+\omega')\,\delta^d(k+k')}
We will self-consistently choose $\Delta$ at the end of the calculation.  We
have
chosen the $\omega^2$ dependence in the denominator for simplicity.  If we had
chosen a
dependence other than quadratic we could always capture the same relative
scaling
of $\omega$ and $k$
by an appropriate choice of $\Delta$ in \euu .

We note an important simplification due to the constraint $\nabla\!\cdot\!
u=0$, \ref\pierre{
We would like to thank P.~Le~Doussal for pointing this out.  See also
P.~Le~Doussal and R.D.~Kamien, {\sl in preparation} (1993).}
namely that there is no difference between taking $u$ to be quenched or
annealed.
The quenched probability distribution for $r$ is just:
\eqn\equen{P_{\hbox{qu}}(r,s) = \int [du]
{\displaystyle{Z_u(r,s)\over \int d^dr\,
Z_u(r,s)}}P[u]}
However, by integrating \erwh\ over space, we find that
\eqn\econserve{{\partial\over \partial s}\int d^dr\,Z_u(r,s) =
\int d^dr\, \nabla\!\cdot\!\left[D\nabla Z_u(r,s) - \lambda u(r)Z_u(r,s)\right]
.}
where we have used the fact that $\nabla\!\cdot\! u=0$.
By gauss's law, the integral on the right hand side will vanish since
$Z_u(r,s)$ will fall to $0$ at infinity.  Thus the normalization of $Z_u$
is $s$-independent.  Since $\int d^d\!r Z_u(r,0)$ is a constant, independent of
$u$, so
is
$\int d^dr Z_u(r,s)$, and it factors out of the
functional integrand in \equen .  But then we see that
\eqn\epann{\eqalign{P_{\hbox{an}}(r,s) &= {\displaystyle{\int [du] Z_u(r,s)
P[u]\over
\int d^dr \int [du] Z_u(r,s) P[u]}}\cr
&= {\displaystyle{\int [du] Z_u(r,s) P[u]\over \int d^dr\, Z_u(r,s) \int [du]
P[u]}}\cr
&= P_{\hbox{qu}}(r,s).\cr}}

Because of the directed nature of the propagator,  it
is possible to argue that $z^* = 2 -\epsilon$ to all orders. Since
this problem can be viewed as quenched disorder, $u$ will not suffer
any nontrivial rescalings, and so $A$ and $B$ will only rescale trivially. In
addition, as we
will argue in Appendix A,  $\lambda$ does not get
renormalized at any order of perturbation theory.

\newsec{Perturbation Theory and the Renormalization Group}

We analyze this model within the context of the dynamical renormalization
group,
along the lines of
\ref\rdrg{D. Forster, D.R. Nelson, and
M.J. Stephen, {Phys.\ Rev.\ A}~{\bf 16} (1977) 732.}.
For a renormalization
group with parameter $\ell$, we rescale lengths according to $k'=e^\ell k$ and
$\omega'=e^{\gamma(\ell)}\omega$, where $\gamma(\ell)$ is an arbitrary function
of $\ell$
to be determined later.  When integrating out high-momentum modes in
a momentum shell $\Lambda e^{-\ell}< k < \Lambda$, we integrate over
{\sl all} values of $\omega$.
We can choose the field $u$ to
have dimension $0$.  We find differential recursion relations:
\eqna\eren{$$\eqalignno{
\der{D} &= D(\ell)\big( -2+z + {\lambda^2 \over 2BD}(1-{1\over
d})A_d\big)&\eren a\cr
\der{\lambda}&= \lambda (\ell)\big( -1+z\big)&\eren b\cr
\der{B} &= B(\ell)\big(d-2\Delta+z\big)&\eren c\cr}$$}
where $z(\ell) = \gamma' (\ell)$ and $A_d = 2(4\pi )^{d/2}/\Gamma (d/2)$
is a geometrical factor.  Putting these together, we can get
a recursion relation for $g=\lambda^2(1-{1\over d})A_d/2BD$ :
\eqn\erena{{dg\over d\ell} = g\left(\epsilon - g\right)}
where $\epsilon = 2\Delta-d$.  Thus in $d\ge 2\Delta$ dimensions there is a
stable fixed point at
$g=0$, whereas if $d<2\Delta$ then there is a stable fixed point at $g
=\epsilon$.
In the details of the calculation, it is essential that $\Delta<2$.  Succinctly
put, this is because the pole in $\omega$ which is used to evaluate the
self-energy correction ($\omega = \sqrt{(B/A)} k^\Delta$) must dominate
the diffusive part of the propagator ($Dk^2$) at small $k$.  We will
check that this constraint holds when finding the self-consistent value of
$\Delta$.

\newsec{Self-Consistent Value of $\Delta$}

Choosing $D$ to be fixed simplifies calculations (though does not change
the results), so we choose $z(\ell)=2-g(\ell)$.  As we show in Appendix A,
at the nontrivial fixed point, $z=2-\epsilon$ {\sl exactly}.  Using this
we find the following two scaling relations for the position and velocity
correlations:
\eqn\epos{\langle\,[r(L)-r(0)]^2\,\rangle = 2dD L^{2/z}}
and
\eqn\evel{\langle\,{dr(L)\over ds}\cdot {dr(0)\over ds}\,\rangle
={4dD(2-z)\over z^2}L^{(2/z)-2}}

Now we would like to look at the tangent field correlations, corresponding
to \epos\ and \evel .  We would like to evaluate
\eqn\eueue{\langle\,u_i(r(s),s)u_i(0,0)\,\rangle =
\int {d^d\!k\over (2\pi)^d} {d\omega\over 2\pi} {e^{ik\cdot r(s)}e^{-i\omega s}
(d-1)\over A\omega^2 + Bk^{2\Delta}}}
Since $\langle r(s) \rangle =0 $, we take $r(s)=0$ in \eueue\ and then check
that
the corrections are themselves consistent.
We find
\eqn\queue{
\langle\,u_i(r(s),s)u_i(0,0)\,\rangle =
{(d-1)A_d\Gamma(1-\epsilon)A^{(d-2\Delta)/2\Delta}\over 2B^{d/2\Delta}\Delta}
L^{{(\Delta-d)/\Delta}}}

Matching the scaling exponents, we have
\eqn\tas{{\Delta-d\over\Delta} = {2-2z\over z} = {4\Delta -2d -2\over
2-2\Delta+d}}
Solving for $\Delta$, we find $\Delta = d/2$ or $\Delta =(d+2)/3$.
The former gives $\epsilon = 2(d/2) -d =0$, and hence $z=2$.  In this
case \evel\ is incorrect, since the second derivative of \epos\ vanishes.
Thus,
we must choose $\Delta=(d+2)/3$.  We note that this solution satisfies both
$\Delta \le 2$ {\sl and} $2\Delta\ge d$ for $d\le 4$, and so we can say that
the critical dimension of this model is $d_c=4$.  Above this, $\epsilon$ is
negative, and we return to the simple random walk fixed point, $z=2$.
Finally, we compute $z=2-2\Delta+d= (d+2)/3$, in complete agreement with
Flory theory.  We point out that the matching of exponents breaks down when
$d=4$, for
if \epos\ is described solely by an exponent, without logarithms, there is no
matching to do -- that is, the exponent in \evel\ would be $0$.  Thus, we do
not
expect, and in fact do not, reproduce the correct logarithmic corrections to
scaling.

If we expand the complex exponent in the integrand of
\eueue\ in powers of $r(s)$ we would find an
expansion of the form
\eqn\pasd{\sum_{j=0}^\infty (-1)^j {(k)^{2j}[r(s)]^{2j}\over (2j)!}}
where we have suppressed the indices.  This sum only contains even powers
since the integration over $k$ will eliminate odd powers of $k$ by rotational
invariance.  Since $\langle\,r(s)^2\,\rangle^j \sim  s^{2j/z}$ and each power
of
$k^2$ will produce a factor of $s^{-2/\Delta}$, we find that the higher order
corrections scale the same way as the leading term if $\Delta=z$, which, in
fact,
it does.  Thus the approximation is truly self-consistent.

\newsec{Conclusions}

One might imagine adding to this model in a number of ways.  One possibility is
to add an explicit self-avoiding term to \erandf .  This would only cause the
same
complications present in Flory theory.  Another possibility would be to add a
small divergence to the field $u$, corresponding to polymer heads and tails
\ref\kln{
R.D.~Kamien, P.~Le~Doussal and D.R.~Nelson, Phys. Rev. A {\bf 45} (1992)
8727.}.  We would
then have
\eqn\enewu{\langle\,u_i(k,\omega)u_j(k',\omega')\,\rangle
= {\delta_{ij} - \alpha{k_ik_j/ k^2}\over A\omega^2 +
Bk^{2\Delta}}\,\delta(\omega+\omega')\,\delta^d(k+k')}
where $\alpha$ is close to, but not equal to $1$.  In this case we find that
new graphs
arise which spoil the exact argument in Appendix A.  Moreover, some of these
new graphs
will diverge logarithmically when $\Delta = \Delta_F = (d+2)/3$.  One might add
a small
correction $\delta$ to $\Delta_F$ and do a $\delta$ expansion around
$\delta=0$.
Unfortunately, now, there is a new fixed point, and the self-consistent
correction $\delta$ is
not small.

It would have been more natural to consider a walk in a random {\sl potential}
and
then self-consistently choose the two-point correlation of the potential to
match the density-
density correlation.  It is, unfortunately, notoriously difficult to study
polymers in
a random potential in an arbitrary dimension \ref\leon{
M.~Kardar and Y.~C.~Zhang, Phys. Rev.  Lett. {\bf
58} (1987) 2087.}.
Here we have exploited the solubility of a random-walker in a
random velocity field.

It is perhaps a curiosity that self-avoidance was not involved in this
calculation.  In fact,
by mapping the system to quantum mechanics, we have actually mapped the system
to
a directed walk in an
external vector field, ignoring the energy cost of self-intersections entirely.
The self-consistent vector field is a strange object, as it depends not only
on the polymer position ${r}(s)$, but also the point along the
polymer $s$.  This is necessary from a geometric point of view, and suppresses
tangent
vector correlations between two monomers far apart along the polymer sequence.
In \zann\ we can imagine integrating out the velocity field $u$.  The
details of this are in Appendix B, and the resulting free energy looks similar
to
that for a self-avoiding walk, with an energy cost for self-intersections.
Reproducing the
Flory exponent then, suggests that the Flory theory may be more robust, and
that
the model studied here is in the same universality class.  The quality
of the Flory prediction may lie in the fact that the self-consistent analysis
here incorporated
an exact result, within the epsilon expansion.

\newsec{Acknowledgements}

It is a pleasure to acknowledge stimulating discussions with L.~Balents,
M.E.~Fisher, M.~Goulian,
P.~Le~Doussal, T.~Lubensky and D.~Nelson as well as a
critical review by P.~Fendley.
This work was supported in part by the National Science Foundation, through
Grant
No.~PHY92--45317, and through the Ambrose Monell Foundation.

\appendix{A}{Exact Value of $z$}
In the above discussion we used the result that $z=2-\epsilon$ at the fixed
point, to all orders
in $\epsilon$ \ref\old{After submission of this manuscript, J.~Bouchaud
informed us of
a similar result in J. Honkonen and E. Karjalainen, Phys. Lett. A {\bf 129}
(1988) 333.}.  This result is similar to that in \rdrg\ where Galilean
invariance
assured that the scaling field of the interaction would only rescale by its
na\"\i ve
dimension.  Our argument will be perturbative.  The first requirement is that
$u$ only rescales by its na\"\i ve dimension.  Since the quenched and annealed
problems are the same, we can regard $u$ as a quenched random field.
While it is typical to absorb the coupling $\lambda$
into $u$, we will not do so here, and hence
it will not be non-trivially renormalized.  We also note that \erwh\ is linear
in
$\psi$ so any nontrivial rescaling of $\psi$ will not affect $\lambda$.

We consider a general graph with an incoming
$\psi$ line, carrying momentum $p$, and outgoing $\psi^*$ line, carrying
momentum
$p'$, and
an outgoing $u$ line, carrying momentum $p-p'$.
The incoming $\psi$ line must
first meet a vertex with $u$.  At this vertex, let $u$ carry away momentum
$k_1$.  The contribution to the graph from this vertex is then proportional to:
\eqn\eAi{p_\mu \left((k_1)^2\delta^{\mu\rho} -k_1^\mu k_1^\rho\right)
= p_\mu\left((k_1)^2\delta^{\mu\rho} - k_1^\mu k_1^\rho\right)}
Thus the graph will be explicitly
proportional to $p^\mu$.
Similarly, the outgoing $\psi^*$ line will emerge from such a vertex.  If the
$u$ line
carries momentum $k_2$, then it will contribute a term proportional to:
\eqn\eAii{\left(p'-k_2\right)^\nu \left((k_2)^2\delta^{\nu\sigma} - k_2^\nu
k_2^\sigma\right)
= (p')^\nu\left((k_2)^2\delta^{\nu\sigma} - k_2^\nu k_2^\sigma\right)}
due to the transverse nature of the $u$ propagator.
Thus the graph will generate a term with {\sl at least} two powers of the
external momentum.
This will not then renormalize $\lambda$ since it is the coefficient of a term
with only
one power of momentum.  Indeed, this graph will generate terms, which by power
counting, are
irrelevant operators.  Hence, the recursion relation for $\lambda$ will simply
be
\eqn\eAiii{\der{\lambda} = \lambda(-1+z)}
to all orders.

Now,
\eqn\eAA{\der{B}=B(d-2\Delta+z)}
and if $D$ has the recursion relation:
\eqn\edrs{\der{D}= D \big( -2 + z + f(g) \big)}
where $f(g)$ represents the perturbation expansion renormalizing $D$,
then $g$ must have the recursion relation:
\eqn\elba{\eqalign{\der{g}
&= g \left[2\{-1+z\} -\{d-2\Delta+z\} - \{-2+z + f(g)\}\right]\cr &
=g[2\Delta - d -f(g)]\cr
&=g[\epsilon - f(g)]\cr}}
since the dimension of $g\sim \lambda^2/BD$ will include a contribution from
$B$ and $D$.
Thus, if there is a fixed point, then $f(g)=\epsilon$ to all orders in $g$.
Hence,
we see that at the fixed point $z=2-\epsilon$ to {\sl all} orders.

\appendix{B}{Effective Self-Avoidance Interaction}

Starting with \zann , the coupling of the polymer to the random field can be
written
\eqn\ecoup{F_{\hbox{int}}=
{\lambda\over 2D}\int d^d\!xds\,\delta^d(r(s)-x)u(x,s)\cdot{dr(s)\over ds}}
Upon integrating out $u$, we find to order $\lambda^2$ (note that this will not
include
the ${\cal O}(\lambda^2)$ term in \erandf )
\eqn\ecoupp{\eqalign{
F_{\hbox{int}}= -\lambda^2&\int {d^d\!k\over (2\pi)^d} ds ds'\, \cr
&{e^{-\sqrt{(B/A)}k^\Delta\vert s-s'\vert}\over 16D\sqrt{AB}k^{\Delta+2}}
\left(k^2\delta_{ij} - k_ik_j\right)e^{ik\cdot[r(s)-r(s')]}{dr_i(s)\over
ds}{dr_j(s')\over
ds'}\cr}}
Because of phase oscillations, the integral over $k$ is small if $r(s)-r(s')$
is large.
If $r(s)-r(s')=0$ we can replace $k^2\delta_{ij}$ by $k_ik_jd$ in the $k$
integration,
through rotational invariance.  Making this substitution then is a good
approximation,
and the corrections are suppressed due to the oscillations.
We then have
\eqn\ecouppp{\eqalign{
F_{\hbox{int}} &\approx -{\lambda^2(d-1)\over 16D\sqrt{AB}}
\int {d^d\!k\over (2\pi)^d} ds ds'\,
{e^{-\sqrt{(B/A)}k^\Delta\vert s-s'\vert}\over k^{\Delta+2}} {d\over ds}{d\over
ds'}
e^{ik\cdot[r(s)-r(s')]}\cr
&= -{\lambda^2(d-1)\over 16D\sqrt{AB}}
\int {d^d\!k\over (2\pi)^d} ds ds'\,{e^{ik\cdot[r(s)-r(s')]}\over k^{\Delta+2}}
{d\over ds}{d\over ds'}
e^{-\sqrt{(B/A)}k^\Delta\vert s-s'\vert} \cr
&= {\lambda^2(d-1)\sqrt{B}\over 16D\sqrt{A^3}}\int {d^d\!k\over (2\pi)^d} ds
ds'\,
k^{\Delta-2} e^{ik\cdot[r(s)-r(s')]} e^{-\sqrt{(B/A)}k^\Delta\vert
s-s'\vert}\cr
&\qquad\qquad\qquad+ \hbox{constant}\cr}}
Within the self-consistent approximation we found $\Delta=(d+2)/3$, so
$\Delta-2 = d-2\Delta =
-\epsilon$.  We now expand in powers of $\vert s-s'\vert$, since by
self-avoidance large
values of $\vert s-s'\vert$ will typically
be accompanied by large values of $r(s)-r(s')$ which will
suppress the integral.  In addition we can expand in powers of $\epsilon$
assuming
that away from the critical dimension $d=4$ we only have a slightly modified
potential, corresponding to a renormalized interaction. The first term
in a double expansion in powers of $\epsilon$ and $\vert s-s'\vert$ is
\eqn\eed{F_{\hbox{int}} \sim \lambda^2\int ds ds'\,\delta^d(r(s)-r(s')),}
looking very similar indeed to a self-avoidance term.  We can also see that the
repulsion is proportional to $\lambda^2$, which itself is proportional to
the expansion parameter of the model discussed in this paper.

\footatend\vfill\supereject\immediate\closeout\rfile\writestoppt
\baselineskip=14pt\centerline{{\bf References}}\bigskip{\frenchspacing%
\parindent=20pt\escapechar=` \input refs.tmp\vfill\eject}\nonfrenchspacing

\bye